\documentclass{article}
\usepackage[utf8]{inputenc}
\usepackage{geometry}
\usepackage{amsmath}
\usepackage{amssymb}
\usepackage{setspace}
\usepackage{graphicx}
\usepackage{xcolor}
\usepackage{natbib}
\usepackage{bm}
\usepackage{amsthm}
\usepackage{multirow}
\usepackage{tabularx}
\usepackage{authblk}
\RequirePackage{graphicx,multirow,bigstrut}
\RequirePackage{bm,mathtools}
\usepackage[flushleft]{threeparttable}
\usepackage{url}

\usepackage{breakurl}
\theoremstyle{plain}
\newtheorem{prop}{Proposition}
\theoremstyle{remark}
\newtheorem{assumption}{Assumption}

\DeclareMathOperator*{\argmin}{arg\,min}
\DeclareMathOperator*{\argmax}{arg\,max}
\newcommand{\pmtf}{\text{PM}_{2.5}}
\newcommand{\notwo}{\text{NO}_{2}}
\newcommand{\cov}{\text{cov}}
\newcommand{\bc}{\color{black}}
\newcommand{\ec}{\color{black}}

\title{A Bayesian Gaussian Process for Estimating a Causal Exposure Response Curve in Environmental Epidemiology}

\author[1]{Boyu Ren}
\author[2]{Xiao Wu}
\author[3,4]{Danielle Braun}
\author[5]{Natesh Pillai}
\author[3,*]{Francesca Dominici}
\affil[1]{Psychiatric Biostatistics Laboratory, McLean Hospital}
\affil[2]{Department of Biostatistics, Mailman School of Public Health, Columbia University}
\affil[3]{Department of Biostatistics, Harvard T.H. Chan School of Public Health}
\affil[4]{Department of Data Sciences, Dana Farber Cancer Institute}
\affil[5]{Department of Statistics, Harvard University}
\date{}

\begin{document}

\maketitle

\begin{abstract}
Motivated by environmental policy questions, we address the challenges of estimation, change point detection, and uncertainty quantification of a causal exposure-response function (CERF). Under a potential outcome framework, the CERF describes the relationship between a continuously varying exposure (or treatment) and its causal effect on an outcome.  We propose a new Bayesian approach that relies on a  Gaussian process (GP) model to estimate the CERF nonparametrically. To achieve the desired separation of design and analysis phases, we parametrize the covariance (kernel) function of the GP to mimic matching via a Generalized Propensity Score (GPS). The hyper-parameters as well as the form of the kernel function of the GP are chosen to optimize covariate balance. Our approach achieves automatic uncertainty evaluation of the CERF with high computational efficiency, and enables change point detection through inference on derivatives of the CERF. We provide theoretical results showing the correspondence between our Bayesian GP framework and traditional approaches in causal inference for estimating causal effects of a continuous exposure. We apply the methods to 520,711 ZIP-code-level observations to estimate the causal effect of long-term exposures to $\pmtf$, ozone, and $\notwo$ on all-cause mortality among Medicare enrollees in the US. A computationally efficient implementation of the proposed GP models is provided in the \textit{\textbf{GPCERF}} R package, which is available on CRAN. 
\end{abstract}

\section{Introduction}
A critically important scientific question in social, economic, and health sciences is whether there is a causal relationship between a continuously varying exposure (e.g., exposure to fine particulate matter $\pmtf$) and a health outcome. For example, the Environmental Protection Agency (EPA) in the US relies on the estimated shape of the exposure response functions (ERF) between air pollution exposures and health outcomes obtained from epidemiological studies to decide whether to lower the National Ambient Air Quality Standards (NAAQS).  In the US, epidemiological evidence has found that health risks persist at $\pmtf$ levels below the current NAAQS (which are set to $12~\mu_g \slash m^3$ for annual exposure), thus suggesting that the NAAQS should be lowered \citep{EPA:link,di2017association,di2017air,wei2020causal,shi2021national,ward2021long}. However, the final decision has not been made. Recently, the World Health Organization has lowered the annual air quality guideline value of $\pmtf$ to $5~\mu g\slash m^3$ \citep{WHO}.
In light of this regulatory context, developing statistical approaches to identify the change points in the ERF has enormous consequences for timely regulatory actions. Specifically, if we identify a change point above which there is strong evidence of adverse health effects, this finding may impact a revision to the NAAQS which could provide great health benefits to the public.

Some scientists have argued that epidemiological studies that estimate the ERF via regression approaches do not provide evidence of causality and therefore should be dismissed when making  air pollution regulatory decisions \citep{EPA:yosemite}. \cite{dominici2017best} argued that there are at least three equally important notions of what constitutes evidence of causality in air pollution epidemiology. The first is causality inferred from evidence of biological plausibility. The second is consistency of results across many epidemiological studies and adherence to Bradford Hill causal criteria \citep{hill1965environment}. The third is the use of causal inference methods. Under certain assumptions, causal inference methods are more robust to model misspecification compared to traditional regression approach when adjusting for measured confounders. In addition, many causal inference methods explicitly enforce covariate balance to address potential confounding, making them ideal analytic approaches to isolate causal relationships.

Many of the approaches for causal inference make the simplifying assumption of a binary exposure (or treatment)  \citep{rosenbaum1983central,rubin1996matching,hernan2000marginal,robins2000marginal,bang2005doubly,van2011targeted}. An important goal of these approaches is to recover covariate balance either implicitly or explicitly \citep{zubizarreta2012using,zubizarreta2014matching,zubizarreta2015stable,wang2019minimal}. Early developments on causal exposure-response function (CERF) estimation for a continuous treatment \citep{hirano2004propensity, robins2000marginal, robins1994estimation} focus on weighting approaches and rely on the Generalized Propensity Score (GPS). However, covariate balance could be hard to achieve in these approaches when the GPS model is misspecified\citep{fong2018covariate}. Doubly-robust approaches \citep{kennedy2017nonparametric,colangelo2020double,schulz2020doubly} overcome this challenge and provide asymptotically unbiased estimates of the CERF when either the outcome model or the GPS model are misspecified. Recently, weighting methods that directly optimize covariate balance (e.g., entropy) have been extended to the continuous exposure setting and exhibit promising performance \citep{kallus2019kernel,vegetabile2021nonparametric}.

As a popular alternative framework to weighting in causal inference, matching has not been explored as extensively in the context of continuous exposures and matching approaches that are robust against misspecification of the GPS and of outcome models are scarce in literature. One exception is the work by \citep{wuGPSmatching,wu2020evaluating}. These authors introduced a GPS caliper matching framework to increase robustness and interpretability in both the design and analysis stages, with the ability to assess explicitly covariate balance of the pseudo-population implied by matching.

Under a Bayesian framework, Bayesian additive regression tree (BART) models have been developed to estimate the CERF flexibly and also to capture potential treatment heterogeneity in large datasets \citep{chipman2010bart,hill2011bayesian,hahn2018regularization,hahn2020bayesian}. The models have been extended to continuous exposure \citep{woody2020estimating} and typically adjust for confounding by including relevant covariates and/or a summary of them (e.g., the GPS) in the regression model.

Although the literature on CERF is rich and rapidly evolving, there are several major unresolved challenges. First, because most of these approaches are developed under the frequentist paradigm, assessing the uncertainty can become computationally intensive as it often relies on resampling methods such as bootstrap \citep{wuGPSmatching}. For Bayesian approaches, it has been shown that the uncertainty estimates of the conditional average causal effects in BART-based models are too optimistic in regions with low overlap compared to other nonparametric Bayesian models focused on smooth functions (e.g. Gaussian process), see invited discussions in \cite{hahn2020bayesian} for details. Tree-based approaches also tend to suffer from the curse of dimensionality, which can be alleviated either by using summary scores, ideally balancing scores, instead of the entire list of covariates in the model, or by inducing sparsity for the regression terms used to adjust for confounding \citep{linero2018bayesian}. Second, in air pollution  epidemiology in the context of exposure to the primary air pollutants ($\pmtf$, $NO_2$ and $O_3$),
the central question is to assess  whether there is a low exposure level below which there is no  evidence of an adverse causal effect on health (i.e., safe level) \citep{pope2015health}. This question can be addressed by identifying change points of the CERF, which are defined as jump-points in the first derivatives of the CERF \citep{muller1992change,goldenshluger2006optimal,berhane2008bayesian,cheng2008kernel}. To our knowledge, existing methods (in the context of regression or causal inference) do not provide a straightforward approach for quantifying the evidence and associated uncertainty of the derivatives of CERF. BART-based models generally produce non-smooth estimates of the ERC and extensions of them that allow for built-in smoothness on the regression function are necessary for the estimation of derivatives, which can be achieved through probabilistic decision trees \citep{linero2018bayesian}. However, even if the estimates of derivatives can be constructed, the uncertainty of these estimates would involve heavy computation, rendering inference on derivatives inaccessible and in turn hindering efficient detection of policy relevant change points.

In this paper we introduce a Gaussian process (GP) approach for nonparametric modeling and estimation of the CERF in the context of exposure to pollution and all-cause mortality. GPs \citep{Rasmussen06gaussianprocesses} can be viewed as distributions over real-valued functions and therefore they offer a Bayesian nonparametric framework for inference of highly nonlinear latent functions from observed data. To estimate an ERF that has a causal interpretation, we introduce a novel parameterization of the covariance (kernel) function of the GP that is purposely specified to exactly match units via the GPS. To maintain the desirable property of separation of the design and analysis phase on causal inference approaches \citep{rubin2008objective}, we introduce a tuning approach for the unknown hyper-parameters as well as the form of the kernel function of the GP by optimizing a GP-induced covariate-balance metric. We show via theoretical arguments that our proposed approach is equivalent to an exact matching algorithm proposed by \cite{wuGPSmatching}. Our method overcomes the aforementioned challenges that are present in the previously proposed methods for estimation of CERFs via 1) a non-parametric outcome model for flexible estimation of CERFs; 2) introducing a Bayesian framework for the estimation of CERFs with automatic uncertainty evaluation through posterior inference; 3) straightforward estimation of derivatives of the CERF as well as its uncertainty with closed-form expressions, which in turn promotes reliable and computationally efficient inference regarding potential change points. In addition, we develop and apply a nearest-neighbor (nnGP) approximation \citep{datta2016hierarchical,datta2016nonseparable,banerjee2017high} of GPs to facilitate the computational scalability of the proposed method.

Via a simulation study, we verify that the proposed GP model can accurately recover the population average CERF and its change points from observational studies, even when the GPS model is misspecified. We also compare the performance of our method to several state of the art approaches in causal inference for estimating the CERF. Our methodology and models are motivated by a massive observational study that includes a nationally representative sample of 68.5 million Medicare enrollees with around 570 million person years, which are aggregated into 520,711 ZIP-code-level observations. We use this dataset to estimate the causal effect of $\pmtf$, $O_3$, and $NO_2$ on all-cause mortality.

\section{Medicare dataset}
\label{sec:EDA}
We have assembled a longitudinal cohort of over 68.5 million Medicare beneficiaries (older than 65) during the period 2000-2016 to estimate the shape of the CERF in the context of ambient exposure to air pollution and all-cause mortality.  Within this cohort, a unique beneficiary ID is assigned to each individual to allow tracking over time up to the date of death. Each record also includes the following individual-level data: age, race, sex and eligibility to Medicaid which is a proxy for low socioeconomic status. We rely on highly spatio-temporally (daily and at 1 km $\times$ 1 km grids) resolved ambient $\pmtf$, $\notwo$, and ozone concentrations, estimated and validated from previously published prediction models \citep{di2019assessing,di2019ensemble,requia2020ensemble}. The place of residence of each individual is only available at the ZIP-code-level; therefore, we have used zonal statistics to aggregate air pollution predictions from 1 km $\times$ 1 km grids to ZIP-codes by year. \bc For each ZIP-code, we have computed the annual averages of $\pmtf$ and $\notwo$ concentrations and warm-season (April 1st through September 30th) average for ozone, which are used as exposures and covariates in our data application. We aggregate the individual-level data on death to Zip-code level and obtain the estimates of annual mortality rate per Zip-code. Specifically, denote by $d_{i,t}$ the total number of deaths in ZIP-code $i$ in year $t\in\{2000,2001,\ldots,2016\}$ and $D_{i,t}$ the corresponding total person-year. The estimated mortality rate at year $t$ and ZIP-code $i$ is then $\hat\lambda_{i,t} = d_{i,t}/D_{i,t}$. We use $\hat\lambda_{i,t}$ as the outcome for our data analysis. \ec

Information on neighborhood socioeconomic status (SES) factors and demographics is available from the US Census Bureau at the ZIP-code level or the county level. We select 12 variables from these SES factors and demographics at the ZIP-code level if not specified otherwise, including average body mass index (BMI; county-level), percent of smokers (county-level), race composition, median household income and house value, percent population and households below poverty level, distribution of educational attainment (college, some college, high school, not completed high school), population density, sex composition, percent of owner occupied housing and census region. In addition, we have information on the following six additional yearly ZIP-code-level covariates; average temperature and humidity in summer and winter, percent of population with dual eligibility to both Medicare and Medicaid and average age at enrollment. In summary, for each of the above 18 potential measured confounders we have  520,711 year by ZIP-code observations. The details of the dataset are described in \cite{wu2020evaluating}.

\begin{figure}
    \centering
    \includegraphics[width=0.56\linewidth]{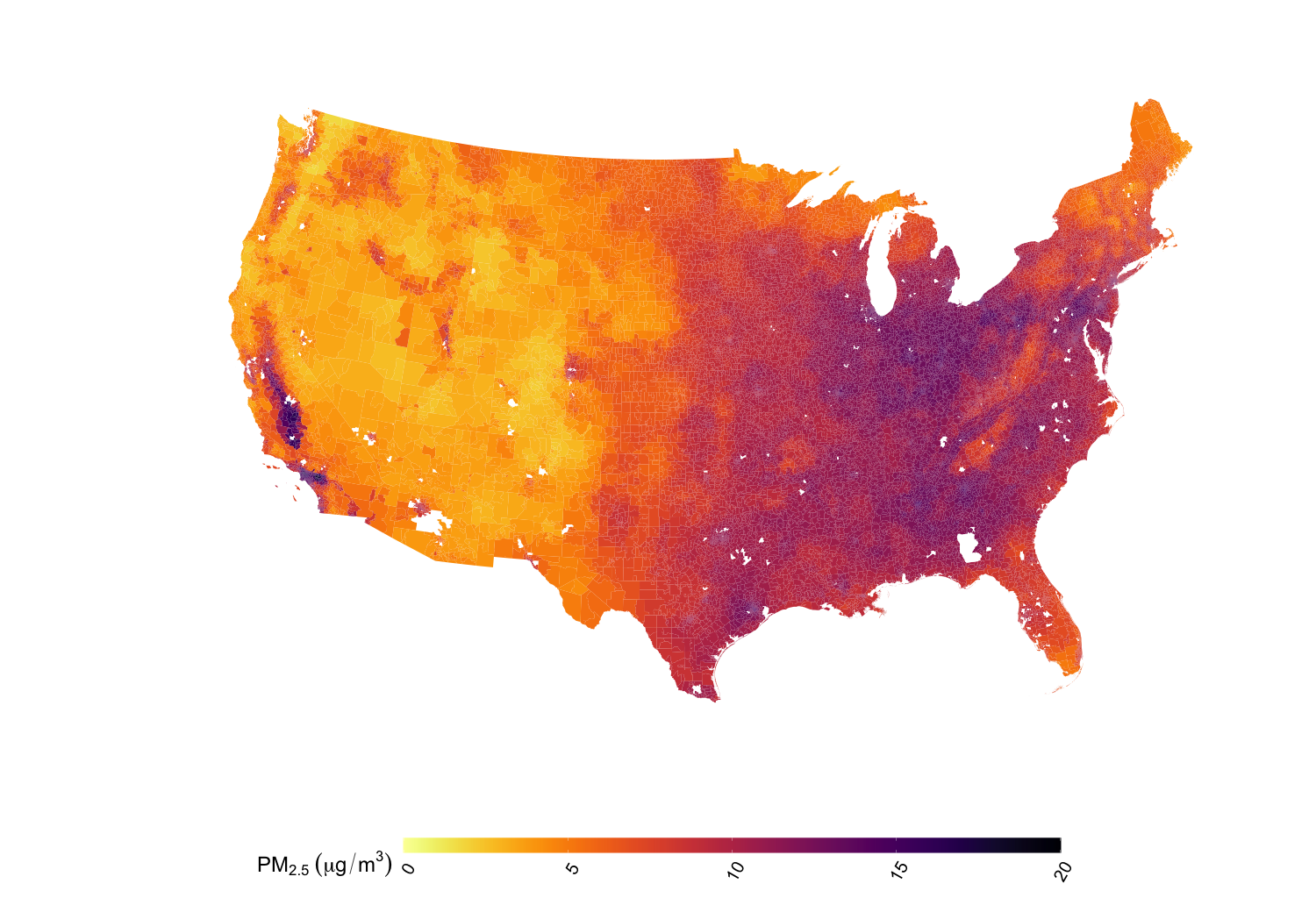}
    \includegraphics[width=0.43\linewidth]{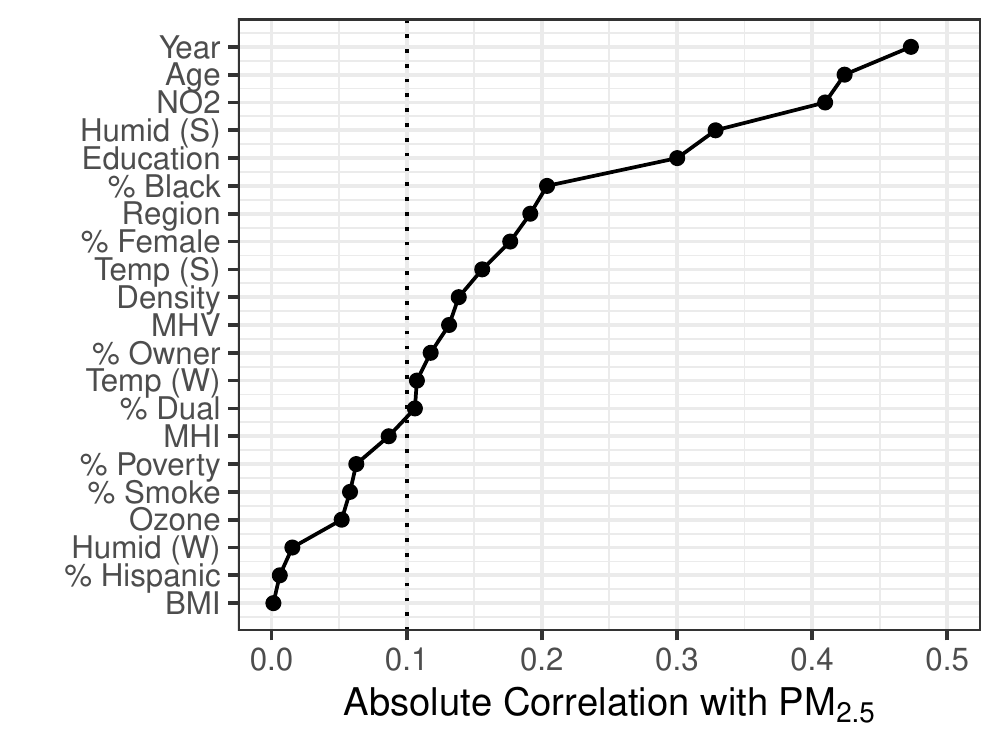}
    \includegraphics[width=0.56\linewidth]{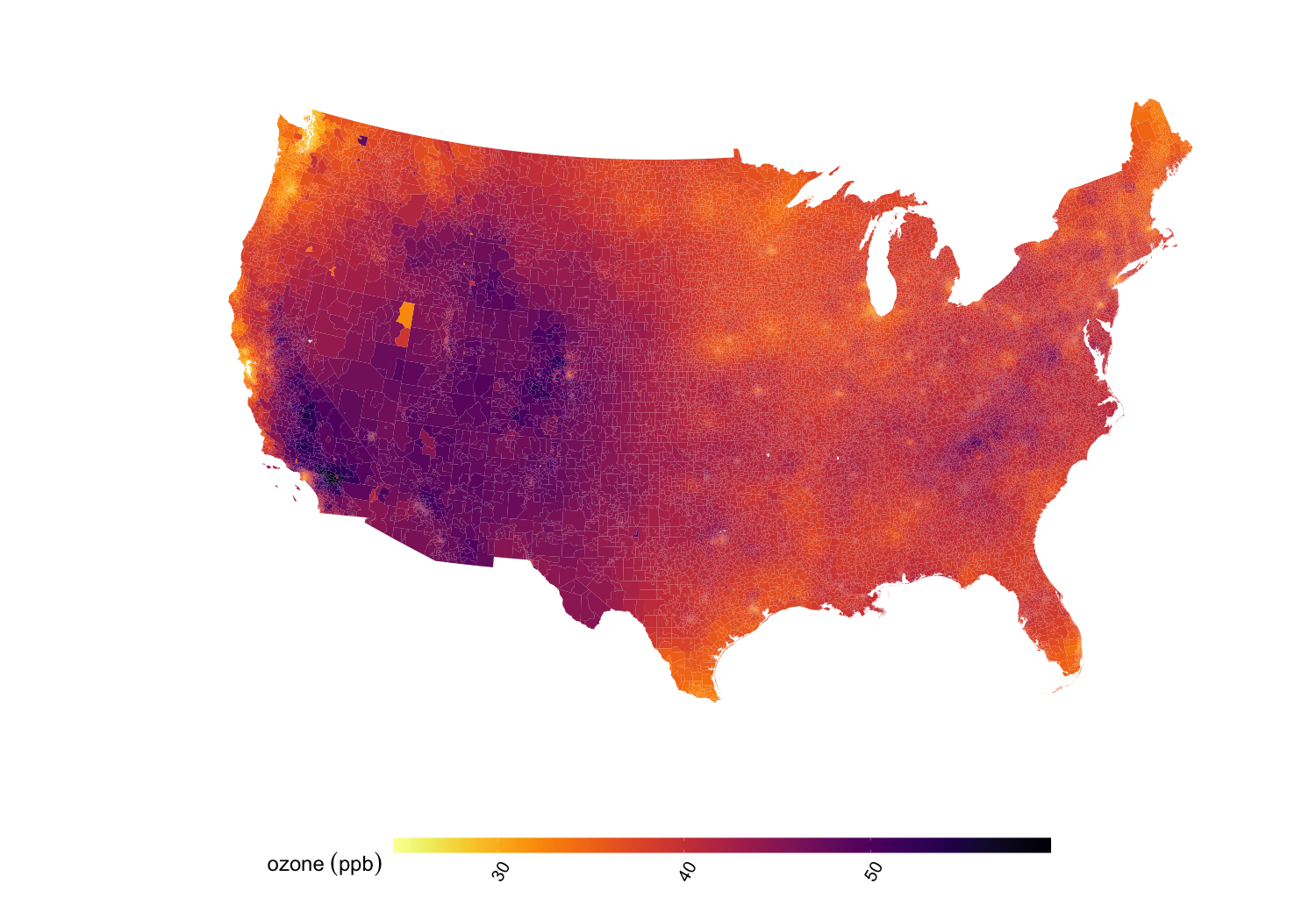}
    \includegraphics[width=0.43\linewidth]{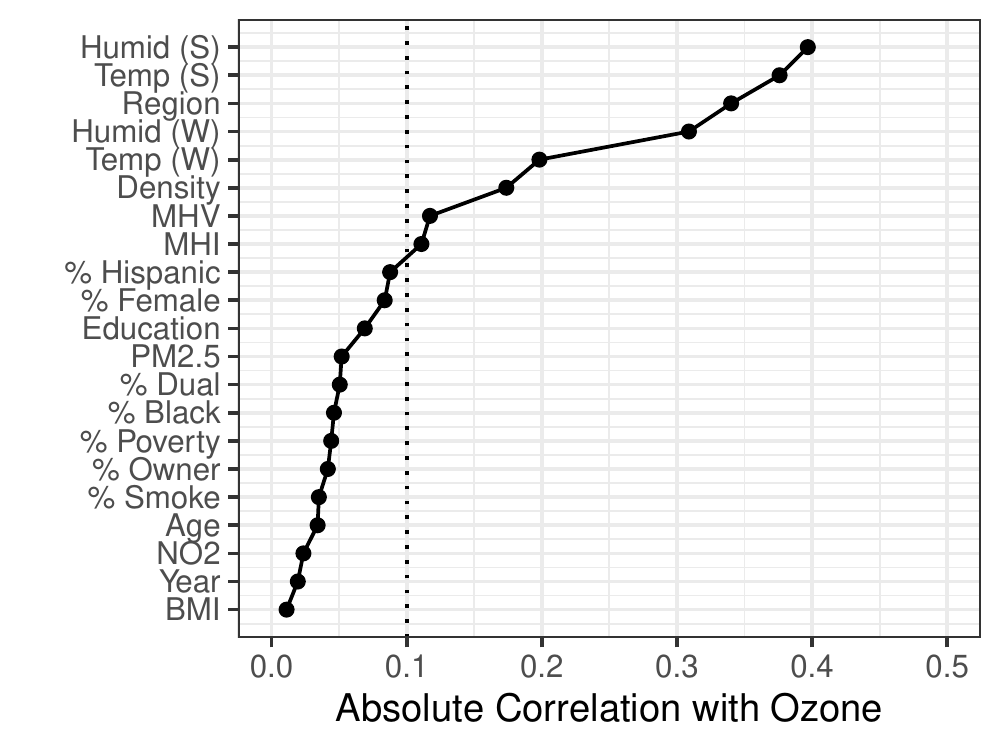}
    \includegraphics[width=0.56\linewidth]{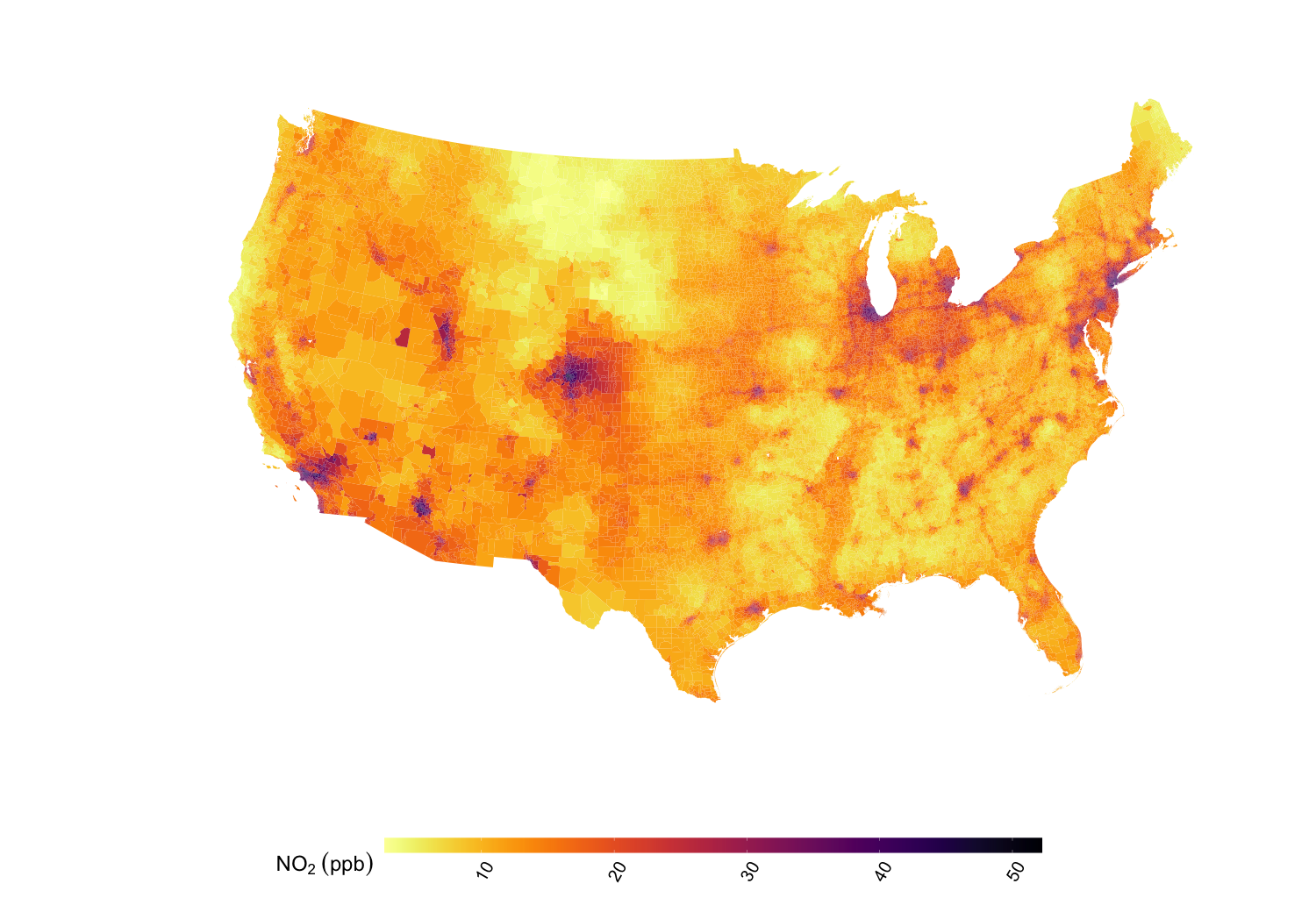}
    \includegraphics[width=0.43\linewidth]{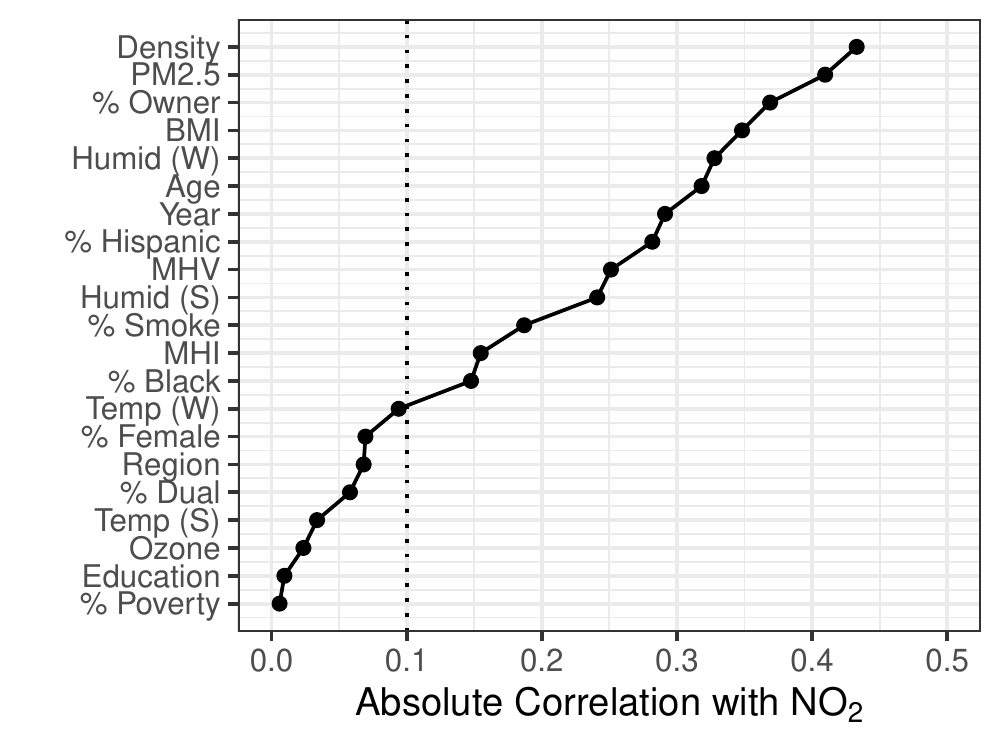}
    
    \caption{\it (\textbf{Left}) ZIP-code-level average concentrations of $\pmtf$, ozone and $\notwo$ across 2000 to 2016. Areas in white are those without $\pmtf$ measurements. (\textbf{Right}) Correlations between each of the three pollutants and other variables. Age $=$ average age at enrollment; Education $=$ proportion of below high school education; Humid (S) and (W) $=$ average summer and winter humidity; Region $=$ census region; Temp (S) and (W) $=$ average summer and winter temperature; Density $=$ population density; \% Owner $=$ \% owner-occupied housing; \% Dual $=$ proportion of dual eligibility of Medicare and Medicaid; MHV $=$ median home value; MHI $=$ median household income; BMI $=$ average body mass index.}
    \label{fig:EDA-res}
\end{figure}

In Figure \ref{fig:EDA-res} left panels, we visualize the ZIP-code-level average concentrations of $\pmtf$, ozone and $\notwo$ averaged across the whole study period (2000 to 2016). We find that $\pmtf$ tends to be higher in the Eastern US whereas ozone is higher in the Western US. $\notwo$ has high concentrations around metropolitan areas. In Figure \ref{fig:EDA-res} right panels, we illustrate the correlations between one of the three pollutants and the other two pollutants as well as 18 ZIP-code-level variables. We calculate these correlations using the ZIP-code level average of the concentrations of pollutants and the 18 variables across 2000 to 2016. Most of the correlations are higher than 0.1, indicating strong covariate imbalance in the dataset. This motivates the need to adjust for these variables when estimating CERFs of the three pollutants. 

\section{GP models for CERF}
\label{sec:method}
 Let $Y_i\in \mathbb R$ be the outcome \bc(for example, mortality rate)\ec, $W_i\in \mathbb R$ be the level of a univariate exposure, and $C_i\in\mathbb R^p$ be the vector of observed covariates for subject $i$. If multiple exposures of interest are present, we will consider one at a time and all the other will be treated as additional confounders. Let $\mathcal S = \{(Y_i,W_i,C_i);i=1,\ldots,N\}$ be the observed data. {\bc We assume that $\mathcal S$ are independent and identically distributed samples from the population.\ec} The random variable $Y_i(w)$ denotes the potential outcome for subject $i$ when the subject is exposed to a level $w$ of a given exposure. Throughout the manuscript, we use capital letters to denote random variables and lower case letters for the realizations of these random variables. As in \cite{hirano2004propensity}, we define the GPS for subject $i$ as $s_i = s(w_i,c_i) = p(W = w_i|C = c_i)$. Under the assumption of consistency (see below),  $Y_i(W_i)$ is observed and equal to $Y_i$. Note that $Y_i(w)$ is unobserved for  $\forall w\neq w_i$. The population average CERF is denoted by $R(w)= \mathbb E(Y_i(w))$, where the expectation is taken over the distribution of the counterfactual outcome in the population of interest. In this section we first introduce the key assumptions that will allow us to identify the expectation of potential outcomes $\left\{Y_i(w)\; \forall w\neq w_i\right\}$ from the observed data $\mathcal S$.

\begin{assumption}
(\textbf{Strong unconfoundedness}) For any $n\in \mathbb N^+$, and $w_1,\ldots,w_n\in \mathcal W= [w^0, w^1]$, the joint distribution of potential outcomes for a subject $i$, $(Y_i(w_1),\ldots, Y_i(w_n))$ is independent of the exposure level $W_i$, conditional on the covariates vector $C_i=c_i$.
\end{assumption} 

\begin{assumption}
(\textbf{Consistency}) The outcome $Y_i$ for subject $i$ which has exposure level $w_i$ is equal to the potential outcome $Y_i(w_i)$.
\end{assumption}

\begin{assumption}
(\textbf{Overlap}) For all values of $c_i$, the conditional density $p(w|C_i = c_i)$ is always positive for any $w\in \mathcal W = [w^0,w^1]$.
\end{assumption}

\begin{assumption}
(\textbf{Smoothness}) The counterfactual outcome $Y_i(w)$ is continuous with respect to the exposure $w$. 
\end{assumption}

Under the strong unconfoundedness assumption, we can factorize the joint density function as
$$
p(Y_i(w), w_i, c_i)=p(Y_i(w)|C_i=c_i) \cdot p(w_i|C_i=c_i)\cdot p(c_i).
$$
We assume a GP model \citep{Rasmussen06gaussianprocesses} for $p(Y_i(w)|C_i = c_i)$:
\begin{equation}
\begin{aligned}
Y_i(w) &= f(w, c_i) + \epsilon_i,\\
f(\cdot) &\sim GP(\mu(\cdot),k(\cdot,\cdot)),\\
\end{aligned}
\label{eq:GP-model}
\end{equation}
where
$\mu(\cdot)$ and $k(\cdot,\cdot)$ are the mean and covariance functions of the GP while $\epsilon_i\sim N(0, \sigma^2)$ is a noise term, capturing the variation of potential outcomes in subjects with identical $w$ and $c$.  In the remainder of this manuscript, we assume $\mu(\cdot) = 0$ if not specified otherwise.

For the rest of this section we assume that all the hyper-parameters, including $\sigma^2$ and the hyper-parameters in $\mu(\cdot)$ and $k(\cdot,\cdot)$, are known and fixed. In Section \ref{sec:tuning}, we will relax this assumption. We model and estimate the GPS,  $p(w|C_i = c)$, using machine learning approaches, such as extreme gradient boosting \citep{chen2016xgboost}, to reduce potential misspecification of the GPS model. As detailed in Section \ref{sec:cov-fun}, the estimated GPS will be incorporated into the specification of the covariance function $k$. \bc Please note that all our analyses will be conditional on the observed set of covariates (excluding the exposure) and therefore we will not specify a model for $p(c)$. \ec

\sloppy Let $\bm Y = (Y_i(w_i);i=1, \ldots, N)^\intercal$, $\bm \mu = (\mu(w_i,c_i);i = 1,\ldots,N)^\intercal$, $\bm \kappa_{i}(w) = (k((w,c_i),(w_j,c_j)));j = 1,\ldots,N)^\intercal$ and $\bm K = (k((w_i,c_i),(w_j,c_j));i,j = 1,\ldots, N)$. Based on the GP model, we have
 $$
\left[\begin{array}{c}
Y_i(w)\\
\bm Y
\end{array}
\right] \sim MVN\left[
\left(
\begin{array}{c}
\mu(w,c_i)\\
\bm \mu
\end{array}
\right), \left(
\begin{array}{cc}
k((w,c_i),(w,c_i))&\bm \kappa^\intercal_i(w)\\
\bm  \kappa_i(w)&\bm K
\end{array}
\right) + \sigma^2 I_{N+1}
\right],
$$
where $I_{N+1}$ is the $(N+1)$-dimensional identity matrix.  The posterior distribution, $Y_i(w)|\bm Y$, follows a normal distribution,
$
N\left(\mu_i(w), \sigma_i^2(w)\right),
$
where
\begin{align*}
\mu_i(w) &= \mu(w,c_i) + \bm \kappa^\intercal_i(w) \left(\bm K+\sigma^2 I_{N}\right)^{-1}(\bm Y - \bm \mu)\\
\sigma_i^2(w) &= k((w,c_i),(w,c_i)) + \sigma^2 - \bm \kappa^\intercal_i(w) \left(\bm K+\sigma^2 I_{N}\right)^{-1}\bm\kappa_i(w).
\end{align*}
Under the assumption of consistency, we have $\bm Y = \bm y = (y_1,\ldots,y_N)$. By strong unconfoundedness, we can estimate $Y_i(w)$
by:
\begin{equation}
\hat Y_i(w) = \mu(w,c_i) + \bm \kappa^\intercal_i(w) \left(\bm K+\sigma^2 I_{N}\right)^{-1}(\bm y - \bm \mu).
\label{eq:est-cf}
\end{equation}
The corresponding estimate of $R(w)$ is
$
\hat R(w) = N^{-1}\sum_{i=1}^N \hat Y_i(w).
$
To quantify the statistical uncertainty of $\hat R(w)$, we first note that the conditional covariance matrix, $\bm \Sigma(w)$, of $\hat{\bm Y}(w) = (\hat Y_1(w),\ldots,\hat Y_N(w))$ given $\mathcal S$ is 
$$
\bm\Sigma(w) = \bm K(w) + \sigma^2 I_N - \tilde{\bm K}(w)^\intercal (\bm K + \sigma^2 I_N)^{-1}\tilde{\bm K}(w),
$$
where $\bm K(w) = [k((w,c_i),(w,c_j));i,j = 1,\ldots,N]$ and $\tilde{\bm K}(w) = [k((w,c_i),(w_j,c_j));i,j = 1,\ldots,N]$. We evaluate the statistical uncertainty of $\hat R(w)$ using the posterior variance 
$
\sigma^2_R(w) = N^{-2}\bm 1_N^\intercal \bm\Sigma(w)\bm 1_N,
$
where $\bm 1_N$ is the all-one vector of length $N$.

When the hyper-parameters are unknown, we need to first select their values for the estimation of $Y_i(w)$. Typically, this can be achieved via fully Bayesian inference by putting hyper-priors on these parameters. \bc However, such approaches are developed in the context of regression, which only produces a correlative measure, rather than the causal relation, between the outcome and the exposure \citep{aronow2016does}. Specifically, the posterior distributions of the hyper-parameters rely on the outcomes variables. As a result, the design stage and the analysis stage are mixed, which could lead to biased estimates of the causal effects \citep{rubin2008objective}. In Section \ref{sec:tuning}, we propose to use a GP-induced covariate balance metric for the selection of hyper-parameters and the kernel function. This method completely separates the two stages by leveraging only the observed exposure and covariates $(w_i,c_i)$ for hyper-parameter tuning. We also provide an R package, \textit{\textbf{GPCERF}}, available on CRAN, to implement the GP models introduced here.\ec

\subsection{Covariance function $k$ for causal inference}
\label{sec:cov-fun}
Our goal is to estimate the CERF, denoted by $R(w)$, using the GP model in (\ref{eq:GP-model}). To adjust for measured confounding bias, we want to estimate $Y_i(w)$ by borrowing information more heavily from units $j$ that are as similar as possible to the unit $i$ with respect to both the exposure level $w$ and the estimated GPS ($\hat{s}_i(w,c_i)$). We achieve this goal by specifying a stationary and non-isotropic covariance function for the GP on the two-dimensional coordinates formed by exposure levels and GPS values:
\begin{equation}
k((w,c),(w',c')) = \gamma^2 h\left(\sqrt{\frac{(\hat s(w,c) - \hat s(w',c'))^2}{\alpha} + \frac{(w - w')^2}{\beta}}\right),
\label{eq:GPS-cov}
\end{equation}
where $\hat s(\cdot,\cdot)$ is the estimated GPS, $h:[0,\infty)\to[0,1]$ is a non-increasing function, the unknown parameters $\alpha$ and $\beta$ control the relative importance of the GPS $\hat s(w',c')$ and the exposure level $w$ when determining the similarity of units, and $\gamma > 0 $ controls the overall scale of the GP. \bc We focus on a distance-based stationary covariance function $k(\cdot,\cdot)$ (i.e., the kernel) as it has a natural connection with matching algorithm (see Proposition \ref{prop:exact-matching}). Of note, there are various choices for the function $h$ that imply different levels of smoothness of the underlying counterfactual outcome $Y(w,c)$. For example, if $h(z) = \exp(-z^2)$ (i.e., Gaussian kernel), then $Y(w,c)$ is infinitely differentiable in the mean-square sense while if $h(z) = \exp(-z)$ (i.e., Mat\'ern-1/2 kernel), $Y(w,c)$ is only continuous but not differentiable in the mean-square sense. We decide to use the data to determine which $h(\cdot)$ is most appropriate for the estimation of CERF. This is motivated by the results in \cite{van2008rates} and \cite{stephenson2022measuring}, which suggest that the performance of a GP model is sensitive to the kernel choice. To this end, we include $h$ as an additional hyper-parameter that will be selected in the hyper-parameter tuning step (see Section \ref{sec:tuning} for details).\ec
 
\subsection{Theoretical Results}
We now show that with the specification of $k$ in (\ref{eq:GPS-cov}), the estimator $\hat R(w)$ indeed identifies the actual CERF $R(w)$. To see this, we start by illustrating a connection between model (\ref{eq:GP-model}) and an exact matching algorithm based on GPS \citep{wuGPSmatching}. We show that the GP model (\ref{eq:GP-model}) is equivalent to exact matching on $(w, \hat s(w,c))$. The exact matching algorithm estimates $Y_i(w)$ with the observed outcome $\tilde Y_i(w) = y_j$ such that $\hat s(w,c_j) = \hat s(w,c_i)$ and $w_j = w$. The existence of subject $j$ is hard to guarantee, especially when $p$ is large. For simplicity, we assume that subject $j$ exists for every $i$ and $w$. The proof is provided in the Supplementary Materials.

\begin{prop}
\label{prop:exact-matching}
Under the model (\ref{eq:GP-model}) with covariance function (\ref{eq:GPS-cov}), if we assume $\mu = \sigma = 0$ and the prior for $\alpha$ and $\beta$ as
$
\alpha \sim \text{Gamma}(\tau_1,1), \beta\sim \text{Gamma}(\tau_2,1),
$
then the estimated $\hat Y_i(w)$ converges \bc point-wise to $\tilde Y_i(w)$ when $\tau_1 \to 0$ and $\tau_2\to 0$ for any $w$ that satisfies Assumption 3\ec.
\end{prop}

When exact matching can be achieved, based on the property of GPS being a balancing function \citep{imbens2000role} and the unconfoundedness assumption, $\hat R(w)$ is an unbiased estimator of the $R(w)$ at any $w$ that satisfies the overlap assumption. In reality, exact matching is usually not feasible and instead of seeking a single observation $j$ to match  exactly with respect to $w$ and $\hat s(w,c_i)$, we identify a set of  observations in a neighborhood of $(w, \hat s(w,c_i))$ for the estimation of $Y(w)$. The covariance function in GP defines a natural neighborhood around each $(w,\hat s(w,c_i))$ by virtue of the fact that posterior mean $\hat Y_i(w)$ is a weighted sum of a small set of $y_j$ with $(w_j, \hat s(w_j,c_j))$ close  to $(w, \hat s(w,c_i))$. The following proposition illustrates that this strategy indeed produces consistent estimates of $R(w)$. See Supplementary Materials for the proof.

\begin{prop}
\label{prop:unbiased}
Let $\mu = 0$, $\alpha>0$, $\beta>0$, $\gamma>0$ \bc and $\sigma>0$ be fixed\ec. Assume that $\hat s(w,c)$ is the true conditional density of $W$ given $C$. If Assumptions 1-3 hold, and in addition $E(Y|W=w,\hat s(w,c)=s)$ is continuously differentiable, then $\forall \epsilon>0$
$$
\lim_{N\to\infty} \Pi\left(|\hat R(w) - R(w)|>\epsilon|\mathcal S\right)  = 0,~a.e.~P_0,
$$
for any $w\in \mathcal W$. Here $\Pi$ indicates the posterior probability and $P_0$ is the joint distribution of $\{Y_i,W_i,C_i\}_{i=1}^\infty$.
\end{prop}

\bc We do not provide results for the convergence rate in the proposition mainly because the rate of the convergence of GP regression is determined by the choice of the kernel function as well as the smoothness of $Y(w,c)$. In particular, \cite{wang2022gaussian} proved that if the smoothness of the GP, as determined by the kernel function is equal to or higher than the smoothness of $Y(w,c)$, an optimal convergence rate can be attained while if the smoothness of the GP is lower than that of $Y(w,c)$, the convergence rate will be smaller than the optimum. \ec

\subsection{Weighting scheme of GP model and covariate balancing}
\label{sec:tuning}
Recall that under model (\ref{eq:GP-model}),
$
\hat Y_i(w) = \mu(w,c_i) + \bm \kappa^\intercal_i(w) \left(\bm K_N+\sigma^2 I_{N}\right)^{-1}(\bm y - \bm \mu).
$
Let $\bm a_i(w) = (a_{i,j}(w);j=1,\ldots, N)^\intercal = \left(\bm K_N+\sigma^2 I_{N}\right)^{-1}\bm \kappa_i(w)$. Note that $\bm a_i(w)$ depends on $\alpha$, $\beta$ and $\gamma/\sigma$. We argue that $\bm a_i(w)$ defines a weighting scheme of all observed subjects for the estimation of $Y_i(w)$. To see this, we fix $\mu\equiv 0$ and it follows that $\hat Y_i(w) = \bm a_i(w)^\intercal \bm y$. When $k(\cdot,\cdot)$ is stationary, $\sum_j a_{i,j}(w)\approx 1$.
In addition if $w$ and $c_i$ are not in the tails of the empirical distributions of $W$ and $C$, then 
any $a_{i,j}(w)$ that is not negligible is positive \citep{Rasmussen06gaussianprocesses}.  This result implies that $\hat Y_i(w)$ is (approximately) a weighted average of all the observed outcomes $\bm y$. Since $\hat R(w) = N^{-1}\sum_i \hat Y_i(w)$, $\hat R(w)$ is also a weighted average of all the observed outcomes in $\bm y$ and the weight for $y_i$ is
$
a_{\cdot,i}(w) = N^{-1}\sum_{j=1}^N a_{j,i}(w).
$

The collection of $a_{\cdot,i}(w)$ can be used to define a covariate balance metric $\rho_r(w)$ for  each covariate $r$, where $r=1,\ldots,p$, which measures the weighted correlation between the $r$-th covariate and the continuous exposure $W$. More specifically we define $$
\hat \rho_r(w) = \sum_i  a_{\cdot,i}(w) w^*_i c_{i,r}^*.
$$
where $w_i^* = \sigma_w^{-1}(w_i - \bar{w})$, $\bar{w} = \sum_i a_{\cdot,i}(w) w_i$ and $\sigma_w = \sqrt{\sum_ia_{\cdot,i}(w)(w_i - \bar w)^2}$.
$
c^*_i = \Sigma_c^{-1/2}(c_i - \bar{ c}),
$
where $\bar{ c} = \sum_i a_{\cdot,i}(w) c_i$ and $\Sigma_c = \sum_ia_{\cdot,i}(w)( c_i - \bar{ c}) (c_i - \bar{ c})^\intercal$. For casual inference, we would like to achieve low absolute correlation between $w$ and every covariate, which can be assessed with $\hat \rho(w) = p^{-1}\sum_{r=1}^p |\hat \rho_r(w)|$. When estimating the population average CERF at a fixed set of $M$ exposure levels $\{w_1,w_2,\ldots,w_M\}$, the overall covariate balance score is
$
\hat \rho = M^{-1}\sum_{m=1}^M \hat \rho_r (w_m).
$
Note that $\hat \rho$ is a function of $\alpha, \beta, \gamma/\sigma$ and $h$.

\bc The metric $\hat \rho$ implies a tuning approach of hyper-parameters using only covariates and exposure but not observed outcome. We select $\alpha, \beta$, $\gamma/\sigma$ and $h(\cdot)$ by minimizing the overall covariate balancing score $\hat \rho$:
$$
\hat \alpha, \hat \beta, \widehat{\left(\frac{\gamma}{\sigma}\right)},\hat h = \argmin_{\alpha, \beta, \gamma/\sigma,h} \hat\rho(\alpha, \beta, \gamma/\sigma, h).
$$
In practice, $\bm a_i(w)$ contains a lot of weights with negligible values and we will set them to be zero exactly (e.g. threshold$=10^{-5}$). Additionally, if $w$ or $c_i$ is far from their observed population averages, $\sum_j a_{i,j}(w)$ will be smaller than 1 by a significant margin. In this case, we will re-normalize $\bm a_{i}(w)$ by $\sum_j a_{i,j}(w)$ after the initial truncation of negligible values. \ec

The estimation of $\sigma$ has to rely on the observed outcomes $\bm y$ as $\sigma$ captures the uncertainty of the outcomes directly. We propose to use the sample variance of the leave-one-out residuals to estimate $\sigma^2$. Let $\hat Y_i(w_i)$ be the estimated outcome for unit $i$ using $y_1,\ldots, y_{i-1}, y_{i+1},\ldots,y_N$ and $e_i = y_i - \hat Y_i(w_i)$ be the residual. We estimate $\sigma^2$ by $\hat \sigma^2 = \sum_ie_i^2/(N-1)$. \bc We argue that this procedure does not violate the requirement of separating design and analysis stages in causal inference since the estimator $\hat R(w)$ does not depend on $\sigma$. \ec

\subsection{Nearest-neighbor GP for scalability}

A limitation of a GP model is its lack of scalability to datasets with large amount of observations due to the excessive computational cost of inversion of covariance matrices. In our data application, there are 520,711 observations and  a standard GP model such as that in (\ref{eq:GP-model}) will be computationally intractable. We overcome this limitation by imposing a sparsity constraint on the covariance matrix via a nearest-neighbor approach \citep{datta2016hierarchical,finley2019efficient}.

To review the idea of the nearest-neighbor GP (nnGP), we consider the joint density of $\bm Y$. For simplicity, we denote $Y_i(w_i)$ as $Y_i$ in the remainder of this section. $p(\bm Y)$ can be decomposed into a series of conditional densities:
$
p(\bm Y) = p(Y_1)\cdot p(Y_2|Y_1)\cdot p(Y_N|Y_1,\ldots,Y_{N-1}).
$
Equation (\ref{eq:GP-model}) implies that
$
Y_i|Y_1=y_1,\ldots,Y_{i-1}=y_{i-1}\sim N(\tilde \mu_i, \tilde \sigma_i^2),
$
where
\begin{gather*}
    \tilde \mu_i = \mu(w_i,c_i) + \bm \kappa_{i,i-1}^{\intercal}(\bm K_{i-1} + \sigma^2 I_{i-1})^{-1}(\bm y_{i-1} - \bm \mu_{i-1}),\\
\tilde \sigma_{i-1}^2 = k((w_i,c_i),(w_i,c_i)) + \sigma^2 - \bm \kappa_{i,i-1}^\intercal (\bm K_{i-1} + \sigma^2I_{i-1})^{-1}\bm \kappa_{i,i-1},
\end{gather*}
where $\bm K_{i} = (k((w_j,c_j),(w_k,c_k));j,k=1,\ldots,i)$, $\bm \kappa_{i,i'} = (k((w_{i},c_{i}),(w_{j},c_{j}));j=1,\ldots,i')^\intercal$, $\bm y_{i} = (y_1,\ldots,y_{i})^\intercal$ and $\bm \mu_{i} = (\mu(w_1,c_1),\ldots,\mu(w_i,c_i))$.

Note that the expressions for $\tilde \mu_i$ and $\tilde \sigma^2_i$ involve the matrix inversion $(\bm K_{i-1}+\sigma^2 I_{i-1})^{-1}$, which is computationally demanding when $i\leq N$ is large. nnGP reduces the computational cost for the estimation of $\tilde \mu_i$ and $\tilde \sigma^2_i$ through a conditional independence assumption. Specifically, denote the top-$\ell$ nearest neighbors of subject $i$ as $\mathcal N_{\ell,i}$, where the distances are determined by the coordinates $(w_i,\hat s(w_i,c_i))$, the nnGP assumes that
$
p(Y_i|Y_1,\ldots,Y_{i-1}) = p(Y_i|\mathcal N_{\ell,i}),
$
which is equivalent to assuming that $Y_i$ and $\{Y_1,\ldots,Y_{i-1}\}\setminus \mathcal N_{\ell,i}$ are independent conditional on $\mathcal N_{\ell,i}$. As a result, the joint distribution of $\bm Y$ can be expressed as a product of multiple low dimensional density functions $p(Y_i|\mathcal N_{\ell, i})$ and posterior inference based on this likelihood function will be much more scalable than a standard GP as $N$ increases. Similarly, for any unobserved point $Y(w,c)$, the nnGP assumes that the conditional distribution $Y(w,c)|\bm Y$ only depends on the top-$\ell$ nearest neighbors of $(w,\hat s(w,c))$ in the observed $N$ samples.

To understand the amount of computational complexity reduced by the nnGP, note that the conditional independence assumption introduced by nnGP makes the precision matrix of any collection of $Y(w,c)$ to be sparse. The inversion of such precision matrix comes with a computational complexity of $O(N\ell^3)$, which is much smaller than the complexity ($>O(N^2)$) associated with computing a precision matrix induced by a standard GP. A typical specification of a nnGP model is described based on an associated standard GP model and requires an additional parameter $\ell$, which indicates the size of the nearest-neighbor sets $\mathcal N_{\ell, i}$.

\subsection{Detecting change points of ERCs}
\label{sec:change-point}
We define the change points of an ERC as the places where a jump in the first derivatives is present. Note that the left and right derivatives of $R(w)$ with respect to $w$ at $w_0$ are
\begin{gather*}
    \frac{\partial_- R(w)}{\partial w}\bigg|_{w = w_0} = \lim_{w\to w_0^-} \frac{R(w) - R(w_0)}{w - w_0},\\
    \frac{\partial_+ R(w)}{\partial w}\bigg|_{w = w_0} = \lim_{w\to w_0^+} \frac{R(w) - R(w_0)}{w - w_0}.
    \end{gather*}
At a change point $w_0$, we have the size of the jump in the first derivatives as
$
\Delta(w_0)\coloneqq\partial_- R(w_0)/\partial w - \partial_+ R(w_0)/\partial w \neq 0.$ We adopt the approach first considered in \cite{muller1992change} and propose an algorithm for change point detection under our Bayesian framework by directly estimating $Delta(w_0)$. We rely on the fact that the derivative of an arbitrary order of a GP (with the restriction imposed by the kernel function) remains a GP itself.

We assume that the kernel function $k$ is at least once differentiable with respect to $w$ in the mean-square sense. Gaussian kernel as well as Mat\'ern-$\nu$ kernel with $\nu>1$ satisfy this requirement. If $Y(w,c)$ is a GP, its derivative $(\partial Y(w,c)/\partial w, Y(w',c'))$ also follows a GP with mean function $(\partial \mu(w,c)/\partial w, \mu(w',c'))$ and covariance function
\begin{equation}
\begin{aligned}
\cov\left(Y(w,c), Y(w',c')\right) &= k\left((w,c),(w',c')\right),\\
\cov\left(Y(w,c), \frac{\partial Y(w',c')}{\partial w'}\right) &= \frac{k((w,c),(w',c'))}{\partial w'},\\
\cov\left(\partial \frac{\partial Y(w,c)}{\partial w},\frac{\partial Y(w',c')}{\partial w'}\right) &= \frac{\partial^2 k((w,c),(w,c'))}{\partial w\partial w'}.
\end{aligned}
\label{eq:GP-deriv}
\end{equation}
\bc Based on the results in (\ref{eq:GP-deriv}), we can leverage the GP device to estimate the one-sided partial derivatives $\partial_-Y(w,c)/\partial w$ and $\partial_+ Y(w,c)/\partial w$ at $w_0$ with the posterior mean of $\partial Y(w,c)/\partial w$ conditioned on all observations with $w<w_0$ and $w>w_0$ respectively. That is,
\begin{align*}
\widehat{\frac{\partial_- Y(w,c)}{\partial w}}\bigg |_{w = w_0} = E\left(\frac{\partial Y(w,c)}{\partial w}\bigg|\mathcal S^-\right), ~~ \widehat{\frac{\partial_+ Y(w,c)}{\partial w}}\bigg |_{w = w_0} = E\left(\frac{\partial Y(w,c)}{\partial w}\bigg|\mathcal S^+\right),
\end{align*}
where $\mathcal S^- = \{(y_i,w_i,c_i)\in\mathcal S: w_i<w_0\}$ and $\mathcal S^+ = \{(y_i,w_i,c_i)\in\mathcal S: w_i>w_0\}$. We then estimate the one-sided derivatives of $R(w)$ as follows.
\begin{equation}
\begin{aligned}
\widehat{\frac{\partial_- R(w)}{\partial w}}\bigg |_{w = w_0} = N^{-1}\sum_i\widehat{\frac{\partial_- Y(w,c_i)}{\partial w}}\bigg |_{w = w_0},\\
\widehat{\frac{\partial_+ R(w)}{\partial w}}\bigg |_{w = w_0} = N^{-1}\sum_i\widehat{\frac{\partial_+ Y(w,c_i)}{\partial w}}\bigg |_{w = w_0}.
\end{aligned}
\label{eq:R-deriv-est}
\end{equation}
\ec

The algorithm of change point detection proceeds as follows.
\begin{enumerate}
    \item \textbf{Estimation of the left and right derivatives of $R(w)$ with respect to $w$}. We apply the results in (\ref{eq:R-deriv-est}) and use only the units with $w\leq w_0$ to estimate the posterior distribution of the left partial derivative at $w_0$. Similarly, the right partial derivative is estimated based on units with $w> w_0$. We then derive the posterior distribution of $\Delta(w_0)$.
    
    \item \textbf{Detection of candidate intervals}. We identify all non-overlapping intervals $\mathcal I$ of $w$ in which the credible intervals of $\Delta(w)$ do not cover zero.
    \item \textbf{Searching for change points}. Within each candidate interval $\mathcal I$, we identify the change point as the maximizer of the posterior mean of $\Delta(w)$: $\tilde w = \argmax_{w\in\mathcal I} |\mathbb E(\Delta(w)|\mathcal S)|$.
\end{enumerate}

\section{Simulation Studies}

We use simulation studies to examine the performance of our approach and compare it with four other approaches in the literature, including GPS adjustment estimator \citep{hirano2004propensity}, inverse probability of treatment weighting (IPTW) estimator \citep{Robins2000}, and the non-parametric doubly robust (DR) estimator \citep{bang2005doubly,kennedy2017nonparametric}, under different model specifications. We also include in our comparison the GPS matching algorithm \citep{wuGPSmatching}. To estimate the GPS, we first use extreme gradient boosting algorithm \citep{chen2016xgboost} to approximate $E(W|C=c)$ and calculate $p(W|C)$ using a normal density with mean $E(W|C=c)$ and variance equal to the residual variance of the gradient boosting fit.

\subsection{Simulation settings}
\label{sec:sim-setting}
We follow the specification in \cite{wuGPSmatching} and assume that $C_i\in\mathbb R^6$ with five continuous components and one categorical component:
$$
(C_{i,1},\ldots,C_{i,4})\sim N(0,I_4),~ C_{i,5}\sim \text{Categorical}\{-2,-1,0,1,2\},~C_{i,6}\sim U(-3,3).
$$
We generate $W_i$ with the following six GPS models as follows.
\begin{enumerate}
    \item $W_i =  9\left[-0.8 + (0.1, 0.1, -0.1, 0.2, 0.1, 0.1)C_i\right] + 17 + N(0, 5)$
    \item $W_i = 15\left[-0.8 + (0.1, 0.1, -0.1, 0.2, 0.1, 0.1)C_i\right] + 22 + T(2)$
    \item  $W_i = 9\left[-0.8 + (0.1, 0.1, -0.1, 0.2, 0.1, 0.1)C_i\right] + 1.5c_{i,3}^2 + 15 + N(0, 5)$
    \item $W_i = \frac{49 \exp\left[-0.8+(0.1, 0.1, -0.1, 0.2, 0.1, 0.1)C_i\right]}{1+\exp\left[-0.8+(0.1,0.1,-0.1,0.2,0.1,0.1)C_i\right]} - 6 + N(0, 5)$
    \item $W_i = 42(1+\exp\left[-0.8+(0.1,0.1,-0.1,0.2,0.1,0.1)C_i\right])^{-1}-18 + N(0, 5)$
    \item $W_i = 7\log\left[-0.8 + (0.1, 0.1, -0.1, 0.2, 0.1, 0.1)C_i\right] + 13 + N(0, 4)$
\end{enumerate}
Note that all six specifications of $W_i$ have the majority of the probability mass in $(0,20)$. We use this range since it is consistent with the range of the observed $\pmtf$ levels in our data application. We generate $y_i$ from an outcome model $[Y_i|W=w_i,C=c_i]\sim N(\mu_i, 10)$, where
\begin{gather*}
\mu_i = -10 - 5*(2, 2, 3, -1, 2, 2)c_i - 5w_i(0.1 - 0.1c_{i,1} + 0.1c_{i,4} + 0.1c_{i,5} + 0.1c_{i,3}^2) + 0.13^2w_i^3
\end{gather*}
\bc The average trend of $Y_i$ over $w_i$ and its variability is specified to mimic those of observed log-mortality rates in the data application. We upscale the outcome $Y_i$ here for clearer visualization.\ec We consider three different sample sizes, $N=200,1000,5000$. For both GP implementations, we use the parameter tuning approach introduced in Section \ref{sec:tuning}. The number of neighbors is $25$ for nnGP. \bc We select $h(\cdot)$ from four candidate kernel functions: Gaussian kernel, Mat\'ern-1/2, Mat\'ern-3/2 and Mat\'ern-5/2.\ec

We evaluate the estimated $\hat R(w)$ based on two metrics, absolute bias and mean-squared error (MSE) to the true CERF $R(w)$. Assume we have $S$ simulation replicates, the two metrics are defined as follows.
\begin{align*}
|\text{Bias}| &= M^{-1}\sum_{m=1}^M |S^{-1}\sum_{s=1}^S \hat R_s(w_{(m)}) - R(w_{(m)})|,\\
\text{MSE} &= (MS)^{-1}\sum_{m=1}^M \sum_{s=1}^S(\hat R_s(w_{(m)}) - R(w_{(m)}))^2,
\end{align*}
where $w_{(1)},\ldots,w_{(M)}$ are equally spaced points from $0$ to $20$ with $M = 200$ and $\hat R_s$ is the estimate of CERF in the $s$-th simulation. To balance the computational cost and stability of the results, we report the $|\text{Bias}|$ and MSE for each approach across $S=1000$ simulate replicates when $N = 200$, $S = 200$ when $N = 1000$ and $S = 100$ when $N = 5000$. Finally, we will examine covariate balance achieved by the GP model using the optimal $\hat\rho$.

\subsection{Simulation results}

We summarize the results of the simulations in Table 1. We find that the absolute bias of the full GP model and GPS matching algorithm are very similar to each other across all six scenarios. nnGP on the other hand, has elevated variance of the estimates of $\hat R$, as indicated by the MSE but the sparsity approximation does not inflate the absolute bias of the Bayesian approach too much.

\begin{table}[htbp]
\begin{threeparttable}
  \caption{\it Comparisons of different CERF estimation methods across simulations. We evaluate the accuracy of the estimates by two metrics, absolute bias and mean squared error (MSE). Scenarios are as defined in Section \ref{sec:sim-setting}.}
    \begin{tabular}{c|c|c|c|c|c|c|c}
    \hline
    GPS   & N     & GP-full & nnGP  & Matching & DR    & Adjusted & IPTW \bigstrut\\
    \hline
    \multirow{3}[2]{*}{Scen 1} & 200   & 4.10 (8.24) & 4.73 (10.21) & 4.96 (24.54) & 1.94 (3.97) & 6.34 (8.51) & 7.42 (9.13) \bigstrut[t]\\
          & 1000  & 2.34 (5.91) & 2.66 (6.77) & 2.73 (13.47) & 1.80 (2.51) & 6.12 (6.93) & 7.28 (7.73) \\
          & 5000  & 1.68 (4.44) & 2.15 (5.26) & 2.08 (6.70) & 1.66 (2.10) & 5.94 (6.54) & 7.39 (7.50) \bigstrut[b]\\
    \hline
    \multirow{3}[2]{*}{Scen 2} & 200   & 12.09 (15.96) & 12.69 (17.83) & 14.14 (30.77) & 45.66 (101.76) & 34.68 (40.61) & 24.78 (40.61) \bigstrut[t]\\
          & 1000  & 10.00 (14.98) & 11.42 (16.53) & 10.32 (20.52) & 42.75 (87.32) & 34.78 (40.01) & 20.45 (35.64) \\
          & 5000  & 6.70 (14.33) & 7.32 (14.87) & 4.24 (13.94) & 40.91 (89.54) & 34.73 (40.78) & 18.44 (33.68) \bigstrut[b]\\
    \hline
    \multirow{3}[2]{*}{Scen 3} & 200   & 4.95 (8.58) & 5.12 (9.53) & 5.76 (23.59) & 2.14 (4.24) & 7.84 (9.42) & 8.72 (10.22) \bigstrut[t]\\
          & 1000  & 3.23 (6.41) & 3.56 (6.89) & 3.53 (13.32) & 1.96 (3.03) & 6.93 (8.45) & 8.78 (9.17) \\
          & 5000  & 2.20 (4.68) & 2.89 (5.01) & 2.55 (7.11) & 2.21 (3.16) & 6.32 (8.30) & 8.78 (8.87) \bigstrut[b]\\
    \hline
    \multirow{3}[2]{*}{Scen 4} & 200   & 4.36 (8.83) & 4.98 (9.77) & 5.68 (22.48) & 1.68 (7.93) & 7.84 (8.35) & 8.05 (9.64) \bigstrut[t]\\
          & 1000  & 2.51 (6.36) & 3.46 (7.72) & 3.12 (13.73) & 1.68 (2.42) & 7.53 (8.03) & 7.90 (8.31) \\
          & 5000  & 1.72 (4.50) & 1.81 (4.93) & 2.07 (7.03) & 1.56 (2.03) & 6.89 (7.53) & 7.93 (8.02) \bigstrut[b]\\
    \hline
    \multirow{3}[2]{*}{Scen 5} & 200   & 3.41 (7.66) & 4.02 (8.36) & 4.46 (20.71) & 2.30 (4.10) & 8.32 (9.78) & 8.42 (10.09) \bigstrut[t]\\
          & 1000  & 2.92 (5.66) & 3.31 (6.12) & 2.80 (13.22) & 1.97 (2.68) & 8.12 (8.99) & 8.37 (8.79) \\
          & 5000  & 2.02 (3.88) & 2.13 (4.21) & 1.59 (6.26) & 2.05 (2.43) & 8.07 (8.12) & 8.49 (8.58) \bigstrut[b]\\
    \hline
    \multirow{3}[2]{*}{Scen 6} & 200   & 4.85 (9.19) & 5.71 (10.62) & 6.38 (24.38) & 23.78 (31.22) & 9.63 (10.76) & 10.76 (12.23) \bigstrut[t]\\
          & 1000  & 3.47 (7.52) & 4.17 (8.23) & 4.05 (14.85) & 17.43 (24.98) & 9.04 (10.12) & 11.02 (11.37) \\
          & 5000  & 3.08 (7.06) & 3.21 (7.48) & 1.57 (8.00) & 11.75 (15.08) & 8.67 (9.77) & 11.27 (11.36) \bigstrut[b]\\
    \hline
    \end{tabular}%
    \begin{tablenotes}
      \small
      \item GP-full: the complete GP model; nnGP: nearest-neighbor approximated GP with neighbor size 25; GPS matching: caliper matching in \cite{wuGPSmatching}; DR: doubly-robust estimator in \cite{kennedy2017nonparametric}; Adjustment: including GPS as a covariate in the outcome model; IPTW: inverse probability of treatment weighting estimator.
    \end{tablenotes}
    \end{threeparttable}
  \label{tab:sim-res}%
\end{table}%

If the relationship between $W$ and $C$ is linear and the GPS values are not heavy-tailed (Scenario 1), all approaches have reasonable performance. The doubly robust approach has the best performance in both absolute bias and MSE among all six approaches across the range of $N$ we considered. GP-full has the second lowest absolute bias and MSE, and the nnGP approximation only mildly affects the performance. When the relationship between $W$ and $C$ stays linear but GPS values are heavy-tailed and include extreme values (Scenario 2), GPS adjustment, IPTW and doubly robust estimators all produce very large MSE, and are not able to reduce confounding bias even with large samples. \bc In contrast, our GP models as well as GPS Matching are robust to the extreme GPS values, creating much more stable estimation than the other three estimators. \ec Although the absolute bias and MSE are much larger than in Scenario 1, the GP-based algorithms are able to leverage the incremental information and are not negatively impacted by extreme GPS values.

When the relationship between $W$ and $C$ is non-linear, as depicted in Scenario 3-6, all approaches tend to have higher MSE and absolute bias compared to the first scenario. However, since the error terms are all normally distributed, the extreme behavior of GPS adjustment and doubly robust estimators observed in Scenario 2 is not replicated except for Scenario 6. \bc Doubly robust estimator has the lowest bias and MSE among all estimators in Scenario 3-5 when $N = 200$ and $N = 1,000$, and are comparable to GP-based algorithms and GPS matching in these scenarios when $N = 5,000$. In Scenario 6, DR suffers substantially from the potential mis-specifications of the GPS model and has the worse performance among all six approaches. In this scenario, GP-based approaches outperform others when $N = 200$ and $N = 1,000$ while GPS matching achieves the lowest absolute bias and MSE when $N = 5,000$. \ec

\begin{figure}
\centering
\includegraphics[width=0.95\linewidth]{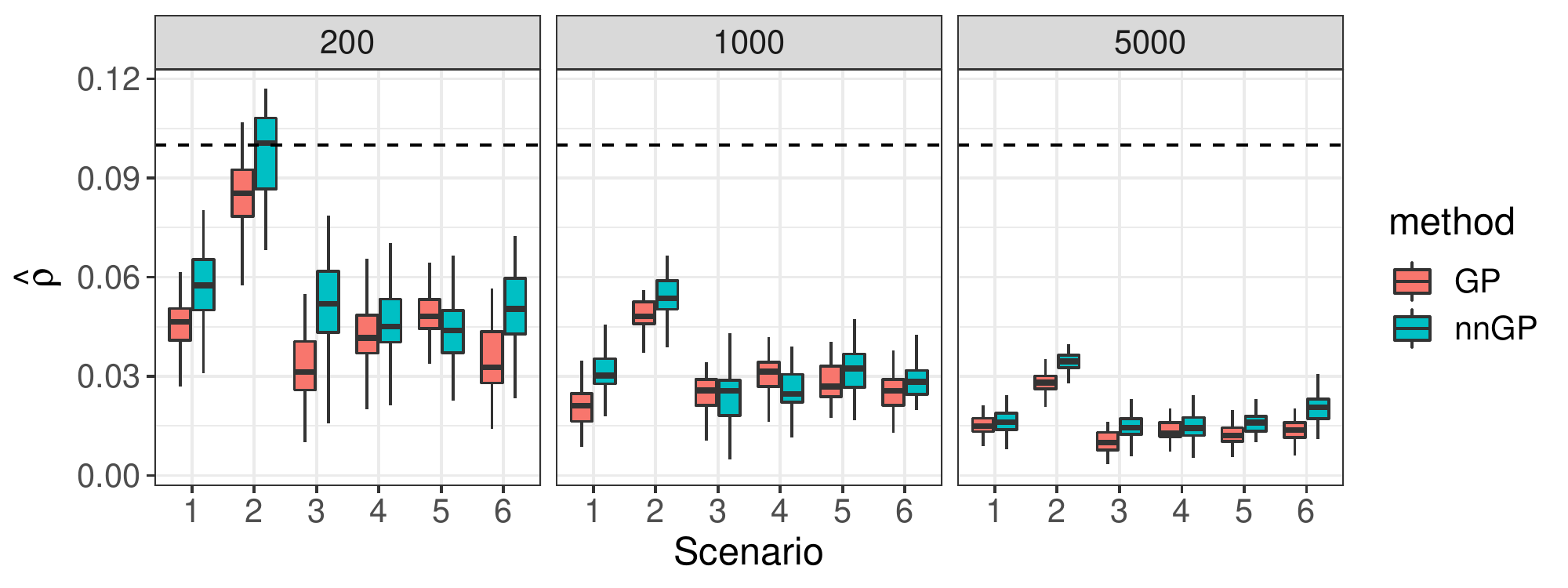}
\caption{\it Covariate balance of GP models in simulation studies. We illustrate the distributions of optimized $\hat\rho$ for the full-GP and nnGP models when the data is generated in each one of the six simulation scenarios. We consider three different sample sizes $N = 200, 1000, 5000$. $\hat\rho$ decreases as $N$ increases, so does its variability. Except for Scenario 2 with $N=200$, all $\hat\rho$ are below 0.1.}
\label{fig:GP-cb}
\end{figure}

\bc We also examine the covariate balance induced by the GP models. In Figure \ref{fig:GP-cb}, we illustrate the distribution of the optimized $\hat\rho$ across simulation replicates for all six scenarios, three sample sizes and two GP models (full GP and nnGP). We can see except for the case where GPS is heavy-tailed (Scenario 2) and sample size is small ($N = 200$), the covariate balance scores are well below 0.1. $\hat\rho$ as well as its variability across replicates decrease as sample size increases. At moderate sample size $N=1000$, $\hat\rho$ in Scenario 2 can also be maintained below 0.1.

Finally, in Supplementary Table S1 and S2, we illustrate the absolute bias of the full GP and nnGP with neighbor size 25 in these six simulation scenarios with each of the four candidate kernels, along with the absolute bias for the optimal kernel selected by maximizing $\hat{\rho}$. We can see that in most cases, the optimal kernel in terms of minimizing covariate balance also achieves the lowest absolute bias. Even in the case where the optimal kernel does not attain lowest absolute bias, its bias is very close to the minimum. The results suggest that minimizing covariate balance metric $\hat\rho$ can effectively select the hyper-parameters and the kernel function. \ec

\subsection{Detecting change-point with GP}
In this section, we use simulated data to examine the ability of our GP model to detect change points of the CERF based on the algorithm in Section \ref{sec:change-point}. We simulate data under the assumption that the true CERF is continuous and piece-wise linear. The place where two linear pieces connect is a change point. Specifically, we assume that
\begin{equation}
\begin{aligned}
\mu_i = &-10 - (2, 2, 3, -1, 2, 2)c_i - w_i(0.1 - 0.1c_{i,1} + 0.1c_{i,4} + 0.1c_{i,5} + 0.1c_{i,3}^2) \\
&+ \mathbb I(2.5<w_i<5)(10w_i-25) + 25\mathbb I(5\leq w<10) + \mathbb I(10\leq w_i<12.5)(10w_i-75) \\
&+ \mathbb I(12.5\leq w_i<17.5)(2.5(w_i-12.5)+50) + 62.5\mathbb I( w_i\geq 17.5).
\end{aligned}
\label{eq:sim-cp}
\end{equation}
We simulate $c_i$ as in Section \ref{sec:sim-setting} and we use the first GPS model. \bc We set the sample size $n = 1,000$ and repeat the simulation for $S = 200$ times. \ec

\begin{figure}[htp]
    \centering
    \includegraphics[width=0.95\linewidth]{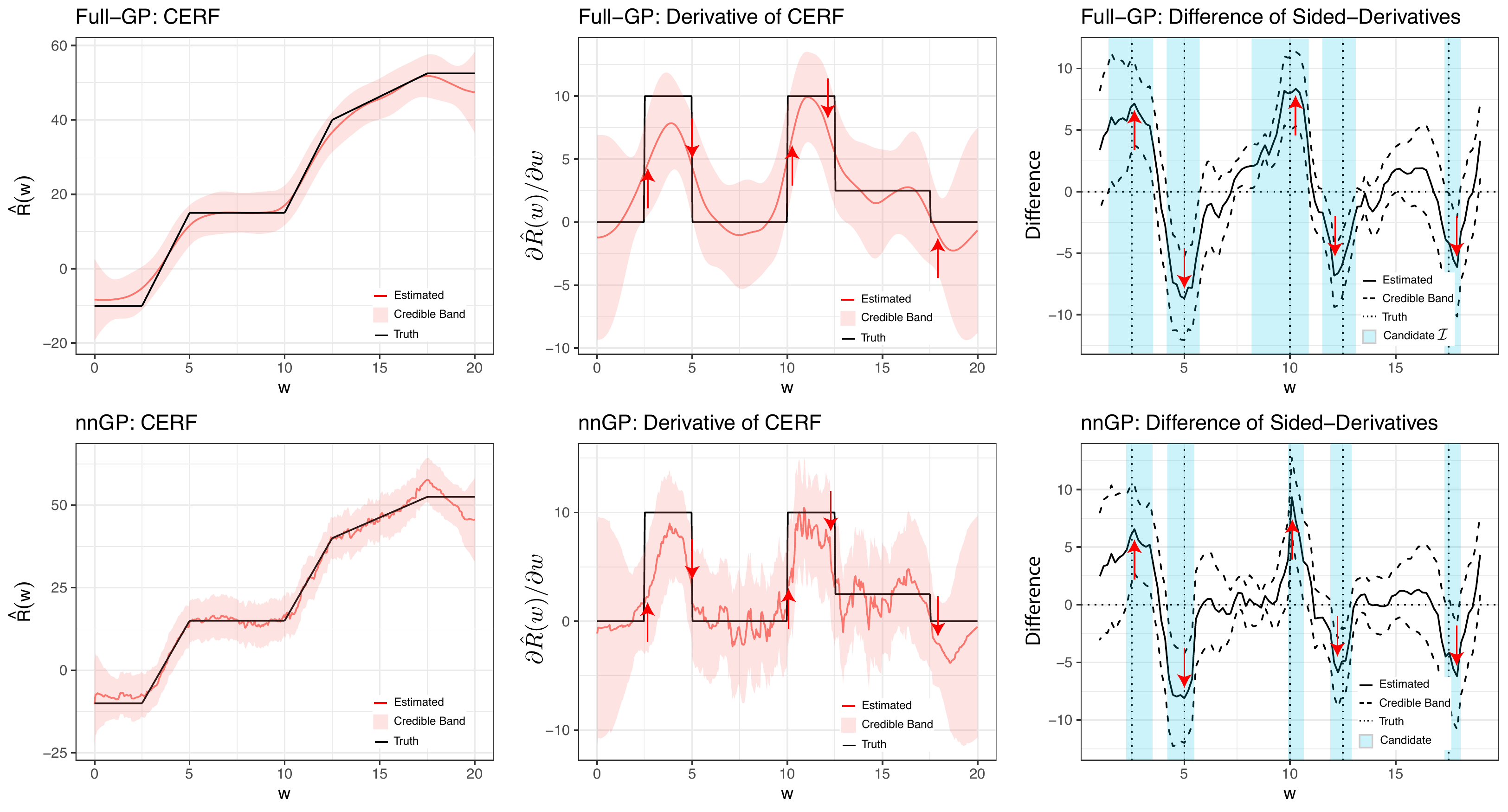}
    \caption{\it Estimating CERF and detecting its change points with full-GP and nnGP models under the piece-wise linear scenario in (\ref{eq:sim-cp}). We illustrate the estimated CERFs (left), estimated derivatives of CERFs (middle) and estimated differences of the one-sided derivatives over the range of $w$ of interest (right). The corresponding credible intervals are also included in the plots. We labeled the candidate regions $\mathcal I$ of change points in the right panels with shaded cyan areas and the detected change points within these regions are marked with red arrows. The direction of the arrows indicates the sign of the jumps in the first derivatives: upward for positive jumps and downward for negative jumps.}
    \label{fig:change_point}
\end{figure}

\bc In Figure \ref{fig:change_point}, we illustrate the estimates of $R(w)$, its first derivative $\partial R(w)/\partial w$ and the results of change point detection (i.e., differences of the two one-sided derivatives with credible bands) under the full GP and nnGP (25 neighbors) models with a Mat\'ern-3/2 kernel in one of the simulation replicates. We find that both approaches recover the true CERF accurately while the estimates from nnGP tend to be less smooth. The estimated first derivative is indeed a smoothed version of the actual derivative (solid lines) and the estimated differences of the sided derivatives have consistent signs and magnitudes with the actual jumps in the first derivatives. Based on the change point detection algorithm in Section \ref{sec:change-point}, we identify the candidate region around each change point (cyan areas in the right panels) and recover all five change points with both full-GP and nnGP, which are labelled with red arrows in the plots. The identified locations of the change point are close to the truth for both full-GP and nnGP. \ec

\bc In Table \ref{tab:cp-detect}, we illustrate a summary of the results for change point detection when we use different kernel functions as well as when the kernel functions are selected by minimizing $\hat\rho$. The results suggest that for full GP, the kernel that matches the smoothness of the underlying $Y(w,c)$ has the best performance for change point detection when no kernel selection is involved. Moreover, with kernel selection added to the analysis pipeline, the performance of the algorithm is optimized on the simulated dataset. For nnGP however, the selected kernel is outperformed by Mat\'ern-3/2 with a small margin. For both GP-related algorithms, selecting kernels will lead to accurate results of change point detection.\ec

\begin{table}[htbp]
  \centering
  \caption{Performance of the change point detection algorithm with different choices of kernel. ``Selection'' corresponds to the analysis pipeline where the kernel for the final estimation is selected by minimizing the covariate balancing score. $C_{p+1/2}$ is the Mat\'ern kernel of order $p$. Note that the true number of change points is 5.}
    \begin{tabular}{c|c|cccc}
    Model & Metric & Exp   & $C_{3/2}$  & $C_{5/2}$  & Selection \bigstrut[b]\\
    \hline
    \multirow{2}[2]{*}{Full-GP} & Average \# of change points & 4.08  & 4.92  & 4.57  & 4.95 \bigstrut[t]\\
          & Average distance from truth & 0.42  & 0.17  & 0.23  & 0.15 \bigstrut[b]\\
    \hline
    \multirow{2}[2]{*}{nnGP} & Average \# of change points & 4.12  & 4.93  & 4.41  & 4.87 \bigstrut[t]\\
          & Average distance from truth & 0.45  & 0.25  & 0.27  & 0.29 \bigstrut[b]\\
    \hline
    \end{tabular}%
  \label{tab:cp-detect}%
\end{table}%

\section{Analysis of Medicare dataset with GP}

We now apply the proposed GP model to Medicare data set described in section \ref{sec:EDA}. Because the place of residence of each subject is only available up to a Zip-code, we aggregate the Medicare data into Zip-code. We first apply the proposed GP model on estimated mortality rate $\hat\lambda_{i,t}$ (520,711 observations) to estimate three separate average CERFs for $\pmtf$, ozone and $\notwo$ on all-cause mortality. We use the log-transformed mortality rate as the outcome in the GP model. The values of the covariates for ZIP-code $i$ and year $t$ are denoted as $c_{i,t}$.  When estimating the causal effect of a pollutant (e.g. $\pmtf$), we also include  the other two pollutants (e.g., ozone and $\notwo$) in the covariate set. We denote the measurement of an air pollutant at ZIP-code $i$ and year $t$ as $w_{i,t}$. For the remainder of this section, we will replace subscript $i$ in Section \ref{sec:method} with $(i,t)$ to emphasize that the unit observation in this dataset is indexed jointly by ZIP-code and year.

We assume that the conditional distribution of $W_{i,t}$ given $C_{i,t}$ is a normal distribution. We estimate the conditional mean of this distribution with SuperLearner \citep{van2007super} and the conditional variance is approximated by the residual variance. We tune the hyper-parameters by optimizing the covariate balance score $\hat \rho$, which is calculated with $M=100$ equidistant levels throughout the exposure range 2.5-17.5 $\mu g/m^3$ for $\pmtf$, 30-50 ppb for ozone and 0-60 ppb for $\notwo$, which are roughly the 0.5 and 99.5 percentiles of the observed levels. \bc Since we are also interested in making inference about change points, we consider three candidate kernels (Gaussian, Mat\'ern-3/2, and Mat\'ern-5/2) to satisfy the requirement that the corresponding GPs are differentiable in the mean-square sense. \ec We use a Gaussian kernel in this application and implement the nnGP approximation to the standard GP model. We set the number of nearest neighbors to be 200. 

\begin{figure}
    \centering
    \includegraphics[width=0.95\linewidth]{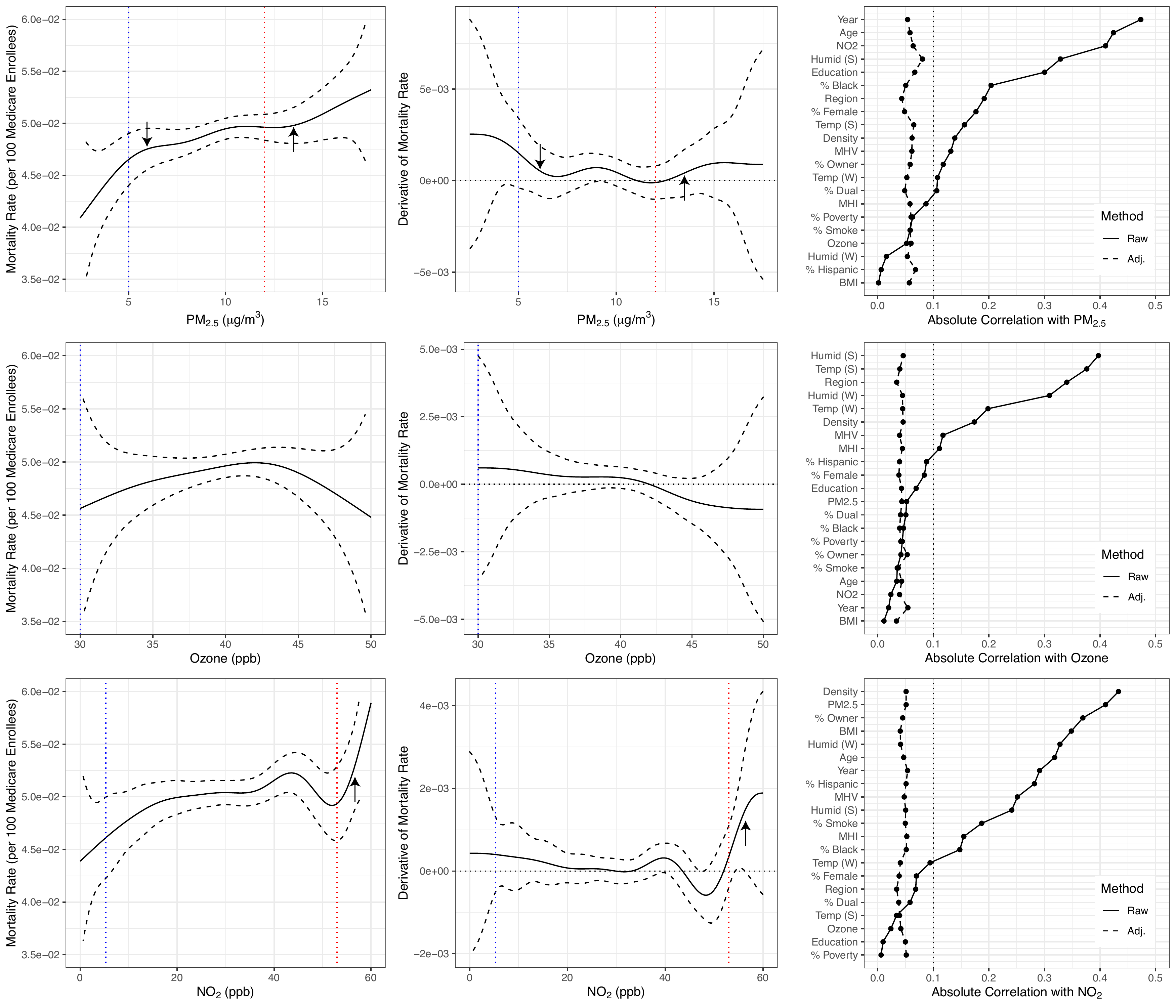}
    \caption{\it Results of the estimated CERFs (\textbf{Left}), their derivatives with respect to the level of an exposure (\textbf{Middle}) and covariate balance in raw and GP adjusted dataset (\textbf{Right}). We consider the curves for $\pmtf$ (\textbf{Top}), ozone (\textbf{Middle}) and $\notwo$ (\textbf{Bottom}). The blue vertical lines indicate WHO standards for all three pollutants ($5 \mu g/m^3$ for $pmtf$, $30$ ppb for ozone and $5.32$ ppb for $\notwo$) and the red vertical lines the EPA standards ($12 \mu g/m^3$ for $pmtf$ and $53$ ppb for $notwo$). Note that the EPA standard for ozone (70 ppb) is not shown in the middle panel since it is larger than the upper limit of the figure. Arrows indicate the detected change points in each of the CERF curves and their directions indicate the sign of the jump in the first derivatives. Age $=$ average age at enrollment; Education $=$ proportion of below high school education; Humid (S) and (W) $=$ average summer and winter humidity; Region $=$ census region; Temp (S) and (W) $=$ average summer and winter temperature; Density $=$ population density; \% Owner $=$ \% owner-occupied housing; \% Dual $=$ proportion of dual eligibility of Medicare and Medicaid; MHV $=$ median home value; MHI $=$ median household income; BMI $=$ average BMI.}
    \label{fig:overall-res}
\end{figure}

Based on the separately selected hyper-parameters and kernel functions, we estimate the three CERFs of $\pmtf$, ozone and $\notwo$ on all-cause mortality. \bc For $\pmtf$ and $\notwo$, the Mat\'ern-3/2 kernel is selected and for ozone, the Gaussian kernel is selected. \ec When estimating the CERF for each pollutant, we exponentiate the posterior means derived from the GP models to recover the estimated all-cause mortality rate $\hat R(w)$. We recover the entire CERF $R(w)$ with the estimates $\hat R(w)$ on 100 equidistant levels of the exposure and calculate the derivative of $\hat R(w)$ with respect to $w$ based on (\ref{eq:GP-deriv}). The statistical uncertainty associated with these estimates are derived using the posterior distribution of the related parameters (Figure \ref{fig:overall-res}). \bc We also visualize the estimated differences of the two one-sided derivatives of $R(w)$ with respect to $w$ along with it credible intervals. We labelled the regions where the credible intervals do not cross zero and indicate the change points according to the algorithm in Section \ref{sec:change-point} (Supplementary Figure S1).\ec

In the left and middle panels of Figure \ref{fig:overall-res}, we illustrate the estimated CERFs $\hat R(w)$ and its derivatives $\partial \hat R(w)/\partial w$ at different levels of a given pollutant while controlling for the other two pollutants as well as the set of demographic and environmental covariates. We also label the current EPA (red) and WHO (blue) standards for all three pollutants with dotted lines. \bc We observe that the mortality rate increases as $\pmtf$ increases. The rate of increase is large when $\pmtf$ is lower than $5\mu g/m^3$ and larger than $15\mu g/m^3$. We detect two change points from the estimated CERFs, one ($14.1 \mu g /m^3$) is close to the current EPA standard ($12 \mu g/m^3$) and the other ($5.94 \mu g/m^3$) is lower than the current standard and closer to the new WHO standard ($5 \mu g/m^3$). The first change point is associated a potential downward change in the derivative, which indicates increase of $\pmtf$ beyond this point tends to cause less increase in mortality rate. On the other hand, the second change point is associated an upward change, which suggests when $\pmtf>14.1$, unit increase in it leads to higher mortality rate. \ec

The estimated CERF for ozone remains relatively constant until 40 ppb, after which a decreasing trend is observed, although the derivative plot suggests this decreasing is not statistically significant. We detect no change points for the CERF of ozone. \bc The estimated CERF of $\notwo$ has a relatively strong increasing trend when the exposure level is lower than 20 ppb and higher than 53 ppb. One change point is detected at 57.4 ppb with a upward change in the derivative. \ec

In the right panels, we illustrate the correlations between each one of the three pollutants and all other variables in the original data and when each observation is weighted according to tuned GPs (weighted pseudo-population). We find that the GP model indeed improves covariate balance and controls the absolute correlation between the exposure and all covariates below 0.1.

\bc We also perform diagnoses on the results to examine whether the model assumptions we make are adequate. Specifically, we examine whether 1) we need to include spatio-temporal components into the kernel $k(\cdot,\cdot)$, 2) we need to consider non-stationary $k(\cdot,\cdot)$ and 3) the over-arching normality assumption of the log-mortality rate is valid. For the first diagnostic task, we estimate the spatial and temporal auto-correlation of the residuals from the GP models. We calculate the Moran's I statistic \citep{moran1948interpretation} with spatial coordinates provided by an approximation of the latitude and longitude of each ZIP-code to examine the level of residual spatial auto-correlation in the GP model. We use the binary weighting scheme as implemented in the R package \textit{lctools} and specify the number of nearest neighbours to be considered as 10. The resulting Moran's I for the model of $\pmtf$ is $6\times 10^{-4}$ with a permutation-based p-value of 0.71, $5.2\times 10^{-3}$ with a p-value of 0.34 for the model of ozone and $7.9\times 10^{-4}$ with a p-value of 0.68 for the model of $\notwo$. For the temporal auto-correlation, we calculate the average absolute auto-correlation with a lag up to 12 years across all zip-codes in the dataset (see Supplementary Figure S2). The results combined suggest that the residuals from the GP fit does not have strong spatio-temporal auto-correlation and capturing potential spatio-temporal variation via fixed effects is adequate for our data.

For the second diagnostic task, we have adopted the visualization approaches in \cite{bastos2009diagnostics} to examine whether this assumption is reasonable. The results (see Supplementary Figure S3) suggest there is no strong evidence against stationary $k(\cdot,\cdot)$, as the standardized residuals of the testing data are nearly symmetrically distributed around zero over the range of the two coordinates (exposure levels and GPS values). For the third diagnostic task, we generate the QQ-plots and histograms of the standardized residuals. We observe the distributions of the standardized residuals have slightly heavier tails than $N(0,1)$ (see Supplementary Figure S4), suggesting the normality assumption is adequate to describe most of the observed data.\ec

\section{Discussion}

We proposed a Bayesian nonparametric model based on GP for the estimation of causal effects of continuous exposures. This model leverages the GPS learned from a flexible machine learning algorithm to adjust for the observed potential confounders. As long as the unconfoundedness assumption holds and the distribution of $W|C$ is correctly specified, the estimated GPS can mirror the actual GPS well. Even if the distribution is misspecified, the estimator of CERFs can still provide valid results when the working distribution of $W|C$ has a similar shape (e.g. normal vs. t-distribution) as the actual distribution. In addition, the posterior inference of the model parameters provides a natural characterization of the uncertainty of these causal estimates. 
\bc We introduced a covariate balance score for the selection of hyper-parameters (including the kernel selection) in the GP model to overcome the mixing of design and analysis stages caused by traditional parameter tuning approaches of GP models.\ec We proved that this Bayesian model is equivalent to an exact matching scheme based on GPS and observed exposure levels and further illustrate that the estimated causal effect derived from our model is consistent, provided the GPS is correctly identified and no unmeasured confounders are present. A unique feature of this model is that the derivatives of a GP process are also a GP. This enabled us to introduce, for the first time, a principled approach to detect change points on the CERF with a full characterization of the statistical uncertainty. A simulation study shows that our model has comparable performance to a recently proposed GPS matching algorithm for the estimation of CERFs \citep{wuGPSmatching} with slight increase in MSE and is robust against certain types of GPS model misspecification. The simulation study also verifies that our proposed algorithm for change point detection has excellent performance.  An application of our model for the estimation of the effects of $\pmtf$, ozone and $\notwo$ on all-cause mortality illustrates that the Bayesian model is scalable when coupled with a nnGP approximation and that our results are consistent with previous findings.

From a methodological perspective, this proposed GP model serves as a foundation for a Bayesian framework that can resolve various issues in the estimation of causal effects of continuous exposures. For example, we can incorporate a model for $p(W,C)$ to take into account measurement error in covariates. From an application perspective, the results on $\pmtf$ reveal potential change points at which a phase transition in the causal effects is observed. It also has relevant policy-making implications: reducing $\pmtf$ further is necessary as the increase of mortality rate associated with increases in $\pmtf$ is most significant when $\pmtf$ is in the low to mid level.

In our model specification, we only correct for the observed covariates by incorporating them into the GPS model. Although in simulations and application we observe that covariate balance can be achieved with this restriction, it might still be beneficial to also include some of the covariates in the main function $\mu$. Ad hoc approaches that select covariates with the highest weighted correlation with the exposure in the weighted pseudo-population imposed by the GP and add them into $\mu$ \citep{rosenbaum2002covariance,abadie2011bias} can be applied here to further improve covariate balance induced by the GP, especially when the GPS model is misspecified, and reduce bias of the CERF estimates.

Throughout the application section, we make the strong assumption that all ZIP-code-by-year observations are independent. We account for the spatio-temporal variation in observations by including year and census region in the GPS model. This assumption can be relaxed by augmenting the two-dimensional index of the GP with time and geographic locations. Another limitation that is worth investigating further is the estimation of GPS. Currently, we use a separate machine learning method to construct the GPS model and use the estimated GPS $s(w,c_i)$ as if it has no statistical uncertainty. As a result, the final characterization of the uncertainty in $\hat R(w)$ is inaccurate. A unified model for $p(Y(w)|C)$ and $p(W|C)$ can be employed to resolve this issue, where a flexible model for $p(W|C)$ can be introduced to mitigate the potential misspecification of the GPS. Bayesian models, for instance those introduced in \cite{giffin2020generalized}, are suitable candidates for this unified approach.

Finally, we want to point out that in the data application we use directly the estimated mortality rate as the outcome and ignore the uncertainty associated with the estimation. Additionally, we exponentiate the estimated log rate to derive the rate in its original scale. These two modeling choices make the posterior credible bands often too optimistic and the posterior mean biased. We would like to extend the GP model in this paper by incorporating a link function and a distribution family of the outcome to directly analyze count data or other types of data that cannot be depicted by a normal distribution. We will focus on the computational aspect of this extended model as the posterior inference would require some form of approximation, likely through variational inference.

%
%

\section*{Acknowledgement}
The authors are grateful to Rachel Nethery for helpful discussions. Funding was provided by the Health Effects Institute (HEI) grant 4953-RFA14-3/16-4, National Institute of Health (NIH) grants R01 GM111339, R01 ES024332, R01 ES026217, R01 ES028033, R01 MD012769, R01 AG066670, R01 MH120400, R33 DA042847, DP2 MD012722 and Gunderson Legacy Fund grant 041622.

\bibliographystyle{apalike} 
\bibliography{reference}       


\end{document}